\documentclass[floatfix, twocolumn]{revtex4}
\usepackage{graphicx, amsmath, amssymb, subfigure}

\begin{document}

\title{Interdependent networks: the fragility of control}

\author{Richard G.~Morris}
\author{Marc Barthelemy}
\email[email: ]{marc.barthelemy@cea.fr}

\affiliation{Institut de Physique Th\'{e}orique, CEA, CNRS-URA 2306, F-91191, 
Gif-sur-Yvette, France}


\begin{abstract}
\textbf
{
	Recent work in the area of interdependent networks has focused on 
	interactions between two systems of the same type.  However, an important 
	and ubiquitous class of systems are those involving monitoring and control, 
	an example of interdependence between processes that are very different.  
	In this Article, we introduce a framework for modelling `distributed 
	supervisory control' in the guise of an electrical network supervised by a 
	distributed system of control devices.  The system is characterised by 
	degrees of freedom salient to real-world systems--- namely, the number of 
	control devices, their inherent reliability, and the topology of the 
	control network.  Surprisingly, the behavior of the system depends 
	crucially on the reliability of control devices.  When devices are 
	completely reliable, cascade sizes are percolation controlled; the number 
	of devices being the relevant parameter.  For unreliable devices, the 
	topology of the control network is important and can dramatically reduce 
	the resilience of the system.
}
\end{abstract}

\maketitle

\begin{figure*}[!t]
	\centering
	\includegraphics[scale = 1.0]{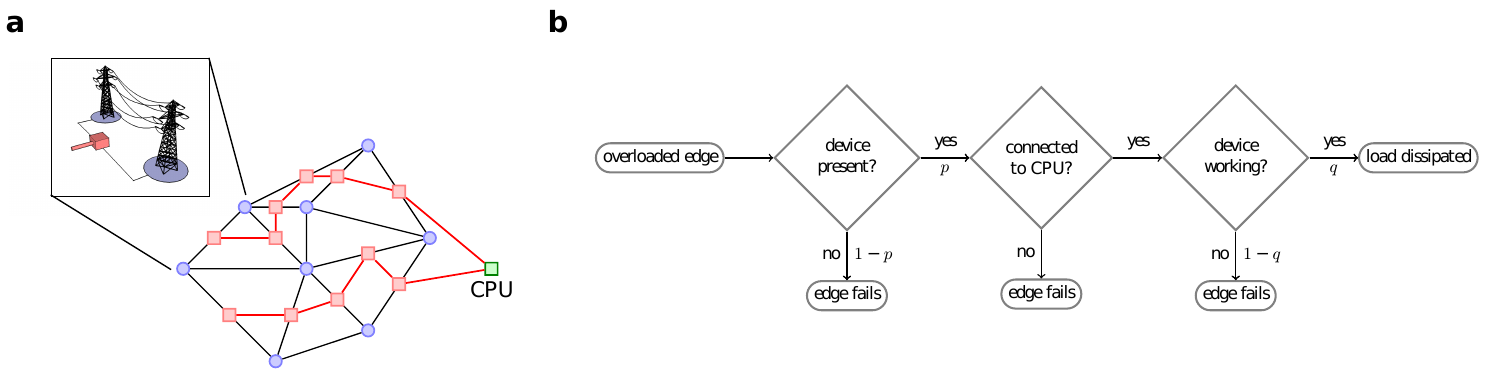}
	\caption
	{
		{\bf Supervisory control.} {\bf a}: The underlying power grid is 
		represented by blue nodes and black edges.  The load carried by 
		power-lines (edges) is supervised by control devices, shown by red 
		squares.  The control devices act as signal relays and form a 
		supervisory network (red edges) with the CPU. {\bf b}: For a given 
		overloaded edge, there is a probability $p$ that a control device is
		present.  If this device is connected to the CPU, then it can be 
		determined if the edge is carrying the largest excess load in the 
		system.  If so, the device will attempt to dissipate the excess load, 
		with a success rate $q$.
	}
	\label{fig:1}
\end{figure*}

The study of interdependent, networked, systems is an area that has recently 
received a lot of attention 
\cite{SVB+10Apr,RP+10Jul,XHJG+11,CG+11Aug,JGSB+11,CDB+12Feb,CDB+12Apr,ASM+12,WL+12,AB+12,RGMMB12},
where the majority of work has so far focussed on the interactions between 
different `critical infrastructures' \cite{MH01,SMR+04,BAC+07,VRLI+08,VV+10}.  
We argue that critical infrastructures should themselves be viewed as a special 
class of interdependent systems, due to the presence of in-built monitoring and 
control mechanisms \cite{SMR+04,DAS04,ANB+08}.  The type of control most 
prevalent in such systems is so-called `supervisory' control--- as 
distinguished from, say, controllability \cite{Liu+11}--- which typically 
involves monitoring an underlying process, with the option of a pre-defined 
intervention once a critical state is reached.  Here, in keeping with the 
picture of interdependent networks, both monitoring and intervention are local 
processes, associated with specific points on the underlying network.  
Furthermore, we are interested in the case when  the control is `distributed', 
that is the local interventions are somehow coordinated via communications 
between sensors.  At the most general level, we are interested in building a 
physics-like model of such systems: that is, complicated enough to encompass 
any interesting behaviour, but sufficiently idealized that the mechanisms at 
play can be easily identified and understood.

Our ideas are based on the supervisory control and data acquisition (SCADA) 
concept, ubiquitous in real-world monitoring of industrial manufacturing, power 
generation, and distribution processes (\textit{e.g.}, electricity, gas, and 
water) \cite{SCADA}.  To this end, our model comprises an underlying system, 
here, an electrical network, where a simple control device is placed on each 
transmission line with a probability $p$ (see Fig.~\ref{fig:1}). The device 
monitors the load of that line and, if it is overloaded, then the device can 
dissipate the excess load with a probability of success $q$, and prevent the 
failure.  In the opposite case, the line fails and the load is redistributed.  
The redistribution of loads may then lead to the overloading and failure of 
further power lines, and so on, potentially resulting in large system-wide 
outages \cite{ID+08}.  If, at any stage during this process, more than one line 
becomes overloaded, then it is assumed that the next line to fail will be the 
one with the largest excess load.  In the case where these lines are 
supervised, it therefore helps if the control devices respond in a coordinated 
way--- always dissipating the excess load on the line under the greatest threat 
of failure.  We therefore stipulate that for a control device to be 
operational, it must be in contact with a central processing unit (CPU).  We 
envisage a communication network composed of ICT-like links connecting the 
devices and the CPU where, in keeping with a distributed SCADA picture, each 
device can also act as a signal relay--- so called `daisy chaining'. Crucially, 
this means that when a control device fails, it can disconnect many other 
devices from the CPU, rendering them useless--- and dramatically increasing the 
fragility of the system. The structure of the supervisory network is therefore 
very important, and we consider two extremes. On one hand, a Euclidean minimum 
spanning tree (EMST) minimises the total length of the control network--- and 
hence the cost--- but typically sacrifices direct connectivity to the CPU.  On 
the other hand, a mono-centric network maximises direct connectivity to the 
CPU, but can be very costly in terms of the total length of network needed.  We 
interpolate between these two configurations by using a simple rewiring 
process: for each node in a EMST, replace with probability $\mu$, the edge 
connected to the neighbor closest to the CPU (along the network), by an edge 
that connects directly to the CPU.  The result is that the topology of the 
supervisory network relies on one continuous parameter $\mu\in [0,1]$, such 
that $\mu=0$ and $\mu=1$ correspond to EMST and mono-centric networks 
respectively.

For modelling the electrical network we adopt a straightforward approach which 
has been proposed and analysed elsewhere \cite{YM+03}. The idea assumes a set 
of producers and consumers linked by power lines, where the resulting load 
carried by each line, or edge, may be represented by a random variable drawn 
from a uniform distribution $U$.  Since $U$ is properly normalized, the upper 
and lower bounds of the distribution are related to the average load $\bar{l}$, 
such that
\begin{equation}
	U(\bar{l}) = \left\{
		\begin{array}{cll}
			1 / 2\bar{l} & \forall\ l\in\left[0,2\bar{l}\right] & \mathrm{if}\ 
			\ \bar{l}\leq 1/2 \\
			1 / 2\left(1-\bar{l}\right) & \forall\ 
			l\in\left[2\bar{l}-1,1\right] & \mathrm{if}\ \ \bar{l} > 1/2
		\end{array}
	\right. .
	\label{eq:Uofl}
\end{equation}
In keeping with the above, it is also assumed that the transmission lines have 
an intrinsic carrying capacity (assumed here, without loss of generality, to be 
one) which, if exceeded, causes the line to fail and the load to be 
redistributed evenly amongst its nearest neighbours \cite{PMB+87}.  The crucial 
departure from Ref.~\cite{YM+03}, is in our choice of network topology.  Since 
many critical infrastructures are, to a good approximation, planar subdivisions 
\cite{MB11}, we use the well known Delaunay triangulation \cite{LGJS85}, which 
is a simple, reasonable model for planar networks such as power grids.

\begin{figure*}[h!]
	\centering
	\includegraphics[scale = 1.0]{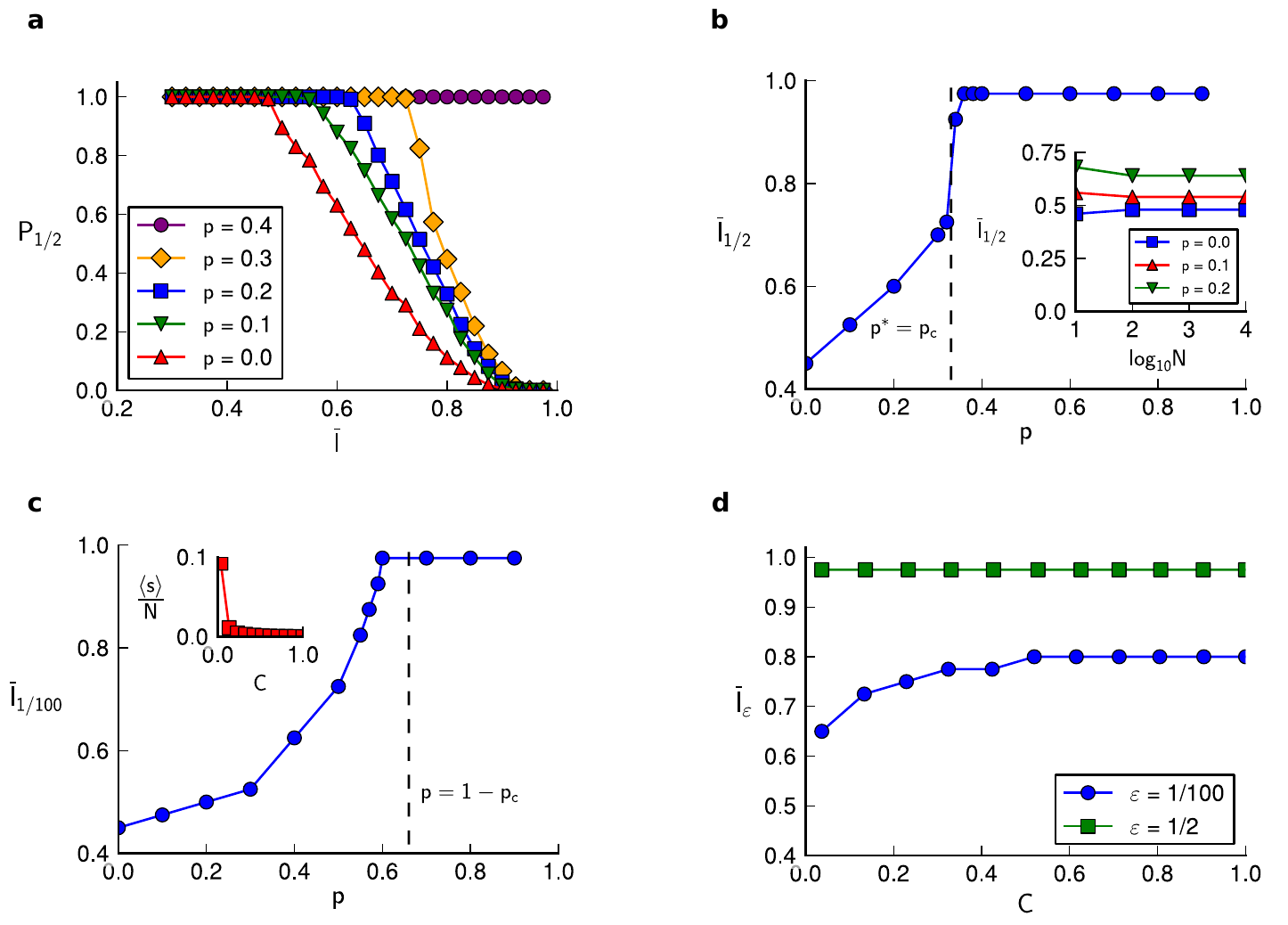}
	\caption
	{ {\bf The effects of reliable control devices ($q=1$).} {\bf a}: The 
	probability that, following a cascade, the remaining largest connected 
	component of the underlying grid contains more than half of the nodes 
	$P_{1/2}$, is dependent on the average load carried by the system 
	$\overline{l}$, and the number of control devices present $p$.  For each 
	value of $p$, the system is characterized by a critical average load 
	$\overline{l}_{1/2}$.  Below this critical value, cascades never disconnect 
	more than half of the system ($P_{1/2} = 1$), whilst above it, there is 
	always a finite chance that this will happen ($P_{1/2} < 1$).  {\bf b}: As 
	the bond-percolation threshold $p^*=p_c\sim0.33$ is approached, the 
	critical value $\overline{l}_{1/2}$ rises sharply to one due to the 
	formation of a giant supervised component (GSC).  Inset: Results are 
	unchanged by increases in system size.  {\bf c}: The bound on cascade size 
	can be lowered by increasing $p>p_c$ and therefore the size of the GSC.  
	For $p\geq 1-p_c$ most nodes are connected by supervised edges and 
	therefore cascades cannot disconnect any nodes completely.  Inset: For 
	values of the cost (estimated as the total length of the supervisory 
	network) above $C\sim 1/2$ the average sub-tree size $\langle s\rangle$ of 
	the control network--- and therefore the average number of devices 
	disconnected at cascade initiation--- is less than 1 and negligible as a 
	fraction of the system size.  {\bf d}: Critical value 
	$\overline{l}_\varepsilon$ for $p>p^*$. In this case, increasing the cost 
	of the supervisory network only increases the critical load associated with 
	bounding small cascades, and not those of the order of the system size.
}
	\label{fig:2}
\end{figure*}

\begin{figure*}[h!]
	\centering
	\includegraphics[scale = 1.0]{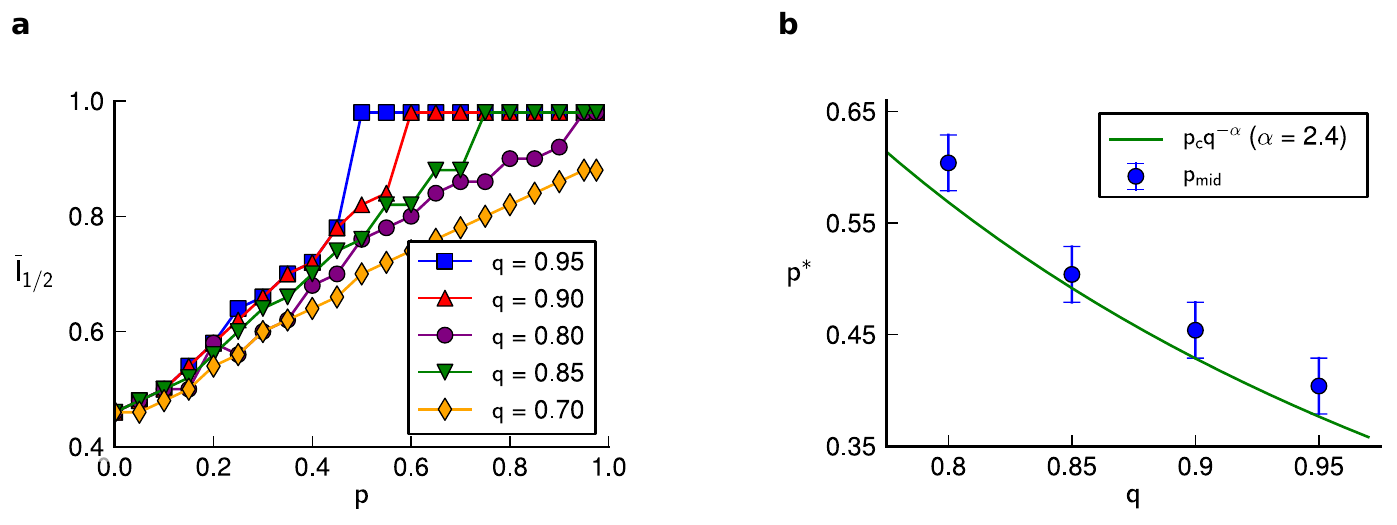}
	\caption
	{
		{\bf The effects of control device failure ($q\neq1$) when every device 
		is connected directly to the CPU ($\mu = 1$).} {\bf a}: As the 
		reliability of the control devices decreases, more devices are needed 
		to maintain the same critical load. {\bf b}: Agreement between the 
		numerical value $p_\mathrm{mid}$ obtained for the $\overline{l}_{1/2}$ 
		transition--- and the theoretical form $p_c q^{-\alpha}$ motivated by 
		simple arguments (see main text and Methods Section).
	}
	\label{fig:3}
\end{figure*}

\section*{Results}

We test the vulnerability of our model against failure cascades by using 
computer simulations (see Methods section for details).  For given values of 
the parameters $p$, $q$, $\mu$, and $\overline{l}$, we repeatedly generate 
instances of the ensemble, each time initiating a cascade according to a 
`fallen tree' approach--- that is, an unspecified external event removes an 
edge and, if it is supervised, the associated control device.  Following each 
cascade, $N_\mathrm{lcc}$, the size of the remaining largest connected 
component of the underlying electricity network, is recorded.  We assume that 
administrators / designers of real systems are interested in ensuring that 
cascades are bounded by a certain size. To this end, we consider
\begin{equation}
	P_\varepsilon = P\left( 1 - N_\mathrm{lcc} / N \leq \varepsilon\right),
\end{equation}
the probability that, following a cascade, the number of nodes disconnected 
from the largest connected component--- the effective cascade `size': 
$1-N_\mathrm{lcc}/N$--- is less than a fraction $\varepsilon\in(0,1]$ of the 
original nodes.

In general, as one would expect, the larger the average load carried by the 
system, the smaller the probability that the cascade size is bounded (see 
Fig.~\ref{fig:2}{\bf a}).  However, we also observe another feature of this 
type of cascading model, first identified in Ref.~\cite{YM+03}: for each value 
of $p$, there is a non-zero critical value
\begin{equation}
	\overline{l}_{\varepsilon} = \sup\left\{\overline{l}\in\left(0,1\right): 
	P_\varepsilon\left(\overline{l}\right) = 1\right\},
	\label{eq:l_epsilon}
\end{equation}
that corresponds to the maximum average load below which cascade sizes are 
bounded with probability one (within a given accuracy, here $1$ part in 
$5\times 10^3$).  Plotting the values of $\overline{l}_{1/2}$ against $p$, a 
sharp transition can be observed at some point $p^*$ (see Fig.~\ref{fig:2}{\bf 
b}).  Above this value, the fraction of disconnected nodes is always bounded by 
$\varepsilon = 1/2$, regardless of how much load the system is carrying.  In 
the completely reliable case ($q=1$), $p^*$ just corresponds to the percolation 
threshold $p_c$ ($\sim0.33$ for Delaunay triangulations \cite{ABRZ09}). The 
cascades are then `percolation controlled' due to the formation of a giant 
component connected by supervised edges, coined here the giant supervised 
component (GSC).  The upper bound on cascade size that is enforced by the GSC, 
can be lowered by employing more control devices--- \textit{i.e.}, increasing 
$p$ (see Fig.~\ref{fig:2}{\bf c}).  For $p\geq 1-p_c$, most nodes are connected 
by supervised edges and cascades cannot disconnect nodes from the giant 
component.

Whilst $q=1$, the only impact of decreasing $\mu$ is to increase the number of 
devices disconnected by the initial external shock.  Disregarding the 
correlation induced by starting the cascade at the point of disconnection, this 
effect corresponds to a small shift
\begin{equation}
	\delta\sim\left\langle s\right\rangle / N,
	\label{eq:delta}
\end{equation}
in the positive $x$-direction of Figs.~\ref{fig:2}{\bf b} and \ref{fig:2}{\bf 
c}.  Here, $\left\langle s\right\rangle$ is the average sub-tree size
associated with a randomly chosen node (see Fig.~\ref{fig:2}{\bf c} inset).  
Figure \ref{fig:2}{\bf d} shows the effects of this shift when $p>p^*$, for 
both large and small $\varepsilon$.  Here, it is natural to characterize 
changes in $\mu$ by a normalized cost
\begin{equation}
	C = L(\mu) / L(1),
\end{equation}
where $L(\mu)$ is the total length of the supervisory network.  The message of 
Fig.~\ref{fig:2}{\bf d} is that: increasing the number of direct CPU 
connections at the cost of increased network length, is only beneficial if the 
suppression of small cascades is desired.

\begin{figure*}[h!]
\centering
	\includegraphics[scale = 1]{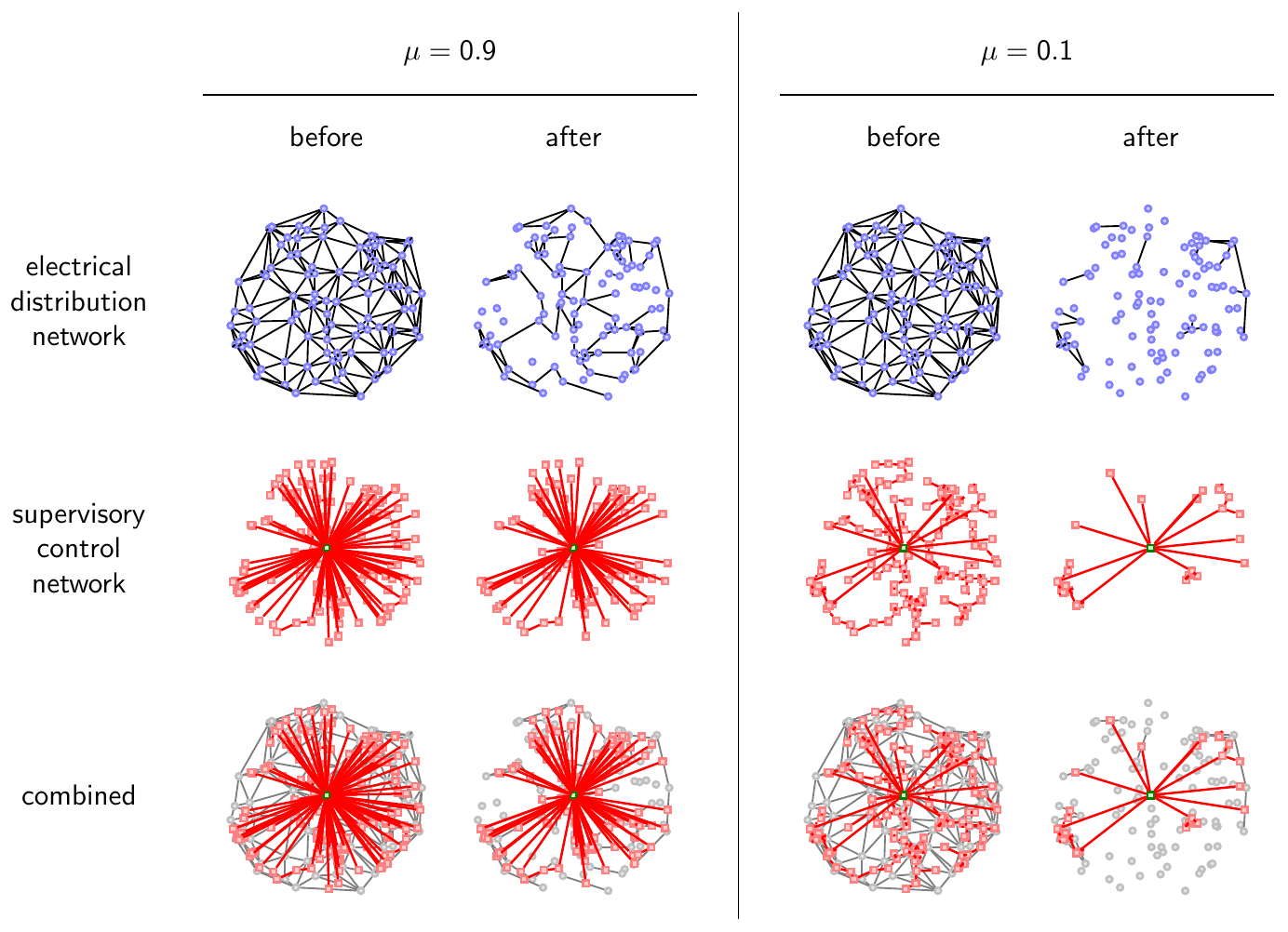}
	\caption
	{
		{\bf The effect of $\mu$ ($p=0.5$, $q=0.9$).}  When the supervising 
		network is almost mono-centric ($\mu=0.9$) very few control devices 
		fail and therefore the remaining largest connected component connects 
		85\% of the nodes in the system.  If the supervising network is 
		almost a tree ($\mu=0.1$) then even though the inherent failure rate is 
		low, many devices become disconnected from the CPU and therefore only 
		10\% of nodes are left connected following a cascade.
	}
	\label{fig:4}
\end{figure*}

\begin{figure*}[t!]
	\centering
	\includegraphics[scale = 1.0]{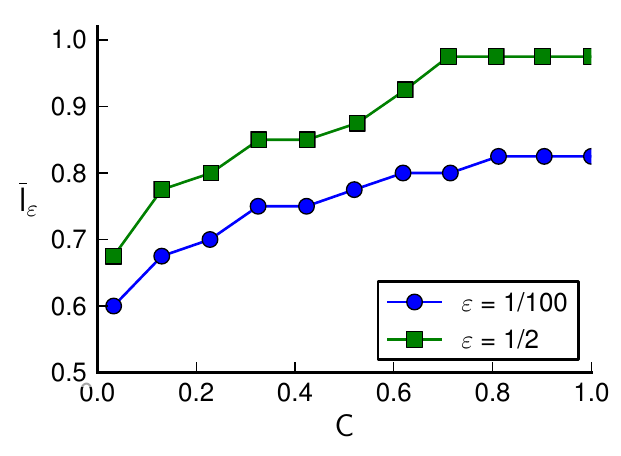}
	\caption
	{ {\bf The effect of topology in control networks with
            unreliable devices ($q=0.9$).} For all cascade sizes
          (ie. regardless of $\varepsilon$), the critical load depends
          strongly the structure of the supervisory network, in
          contrast with the completely reliable case. In particular, even when there
          are already many direct-CPU links ($C>1/2$), the critical
          load that the system can carry is drastically reduced by
          introducing more dependency into the supervisory network.}
	\label{fig:5}
\end{figure*}

If, in contrast to above, the control devices have an inherent rate-of-failure 
($q < 1$), then a GSC may be either disconnected or reduced in size as control 
devices fail.  In the best case scenario, when the supervising network is 
mono-centric and $q$ is close to one, the picture is one of `effective 
percolation' with (see Methods)
\begin{equation}
	p^*=p_cq^{-\alpha},
	\label{eq:pstar}
\end{equation}
where $\alpha$ is determined by the topology of the underlying network ($\sim 
2.4$ for a Delaunay triangulation, see Methods section for details).  This 
simple form shows good agreement with direct estimates of the value of $p^*$ 
(see Fig.~\ref{fig:3}{\bf b} and Methods for details).  For lower values of 
$q$, percolation-like descriptions are no longer appropriate: regardless of the 
number of control devices, it is not possible to bound cascade sizes in a way 
that is independent of the average load carried by the system.  Indeed, if 
control devices are both unreliable ($q < 1$) \textit{and} the control network 
is tree-like ($\mu < 1$), the system is very susceptible to large failure 
cascades, with little impact made by increasing $p$ (see Fig.~\ref{fig:4}).  In 
this case, we see that for both large and small cascades, the topology of the 
control network is very relevant and can induce extreme fragility in the control system (see 
Fig.~\ref{fig:5}).

\section*{Discussion}

In conclusion, we have introduced a minimal model which incorporates
the salient features of many real-world control systems.  Firstly, the
control devices are simple: they only have so-called `supervisory'
functions of monitoring and intervention.  Secondly, the system is
`distributed', that is, not only are the devices positioned in space
but they require coordination--- in this case, by connection to a CPU.
Thirdly, we also incorporate the effects of devices having an inherent
rate-of-failure.  With only these simple characteristics, the
resulting behaviour is very rich. The primary feature concerns the
fragility of such control systems: a small reduction of control device
reliability leads to a regime where the ability to suppress cascades
is dramatically affected by the topology of the control network. Our
results suggest that it is much more cost-effective to try to improve the
reliability of control devices rather than working on the stability of
the supervisory control network. We believe that these results make a
first step in understanding distributed supervisory control, whilst
also providing helpful guidelines to designers and administrators of
real systems.  We welcome further work in the area.

\section*{Methods}

\subsection*{Simulations}

To simulate the system, $N$ nodes are placed in the plane at random, the 
Delaunay triangulation is then formed, and loads are allocated to the resulting 
edges according to $U(\bar{l})$.  The supervisory network is incorporated by 
first adding a control device to each edge with probability $p$, then forming 
the network according to the rewiring procedure described in the main text 
(dependent on parameter $\mu$).  Cascades are initiated by assuming an external 
event that causes an edge to be removed at random and its load is redistributed 
amongst its nearest neighbors.  If the failing edge was supervised, then the control 
device is also removed.  During the ensuing cascade, we stipulate that for a 
control device to work, it must be connected to the CPU, a special node that 
cannot be removed.  If a control device is unconnected, then it cannot work and 
is of no use.  However, if a control device is connected, and it is supervising 
an edge that is about to fail--- \textit{i.e.}, it is carrying the largest 
excess load in the system--- then there is a probability $q$ that the excess 
load is dissipated and the load of the edge is reset to $\bar{l}$.  The 
quantity $q$ can be thought of as the inherent reliability of a
device.

Simulations were written in C++ and implemented using the Boost Graph library 
\cite{boost} where possible.  Delaunay triangulations were produced using an 
iterative algorithm \cite{LGJS85}.

Results are presented for systems of size $N=500$ ($\sim3\times 10^3$ edges) 
and statistics are calculated over $5\times10^3$ instances of each ensemble 
(defined by parameters $p$, $q$, $\mu$, and $\overline{l}$).  Critical values 
$\overline{l}_\varepsilon$ and $p^*$ are accurate up to an error of 
approximately $\pm 0.02$, since they are identified by varying the underlying 
parameter by finite increments.  In Figs.~\ref{fig:4}{\bf c} and \ref{fig:5}, 
$\overline{l}_\varepsilon$ corresponds to $P_\varepsilon > 0.99$ in order to 
accommodate the noise associated with different control network structures.

\subsection*{Formation of an effective GSC}
\begin{table}[!t]
{\small
	\hfill
\begin{tabular}{l|cc} \hline \hline
	& $\langle n\rangle$ & $\mathrm{Var}\left[ n\right]$ \\ \hline
	$q=0.95$,\ $p=0.4$ & 2.36 & 0.0004\\
	$q=0.9$,\ $p=0.45$ & 2.40 & 0.0007\\
	$q=0.85$,\ $p=0.5$ & 2.40 & 0.001\\
	$q=0.8$,\ $p=0.6$ & 2.44 & 0.001\\\hline
\end{tabular}
\hfill}
\label{tab:1}
\caption{For different values of $p$ and $q$, $\langle n \rangle\sim2.4$, and 
	$\mathrm{Var}\left[ n \right] \ll \langle n \rangle$.}
\end{table}
Labelling each supervised edge by $i = 1,\ 2,\ \dots,\ E_s$, the probability 
that a supervised edge survives a cascade is $q^{n_i}$, where $n_i$ is the 
number of times a device is solicited--- \textit{i.e.}, it tries to dissipate 
its excess load with probability $q$.  Here, for large enough systems the 
number of supervised edges is given by $E_s = pE$. (Since the average degree of 
a Delaunay triangulation is peaked around six, the total number of edges $E$ is 
well approximated by $E\sim3N$.)  Using a bar to denote system average 
$\overline{q^{n}}=1/E_s\sum_i q^{n_i}$, we know that if 
$\mathrm{Var}\left[n\right]$ is small, then $\overline{q^{n}}\sim 
q^{\overline{n}}$.  Approximating a large system average with an ensemble 
average $\left\langle\dots\right\rangle$ over many smaller systems, the results 
are given in Table 1.  Here it is clear that the average $\langle n\rangle$ is 
well approximated by the value $2.4$, regardless of $p$ and $q$, and that the 
variance is always very small compared to the average.  We can then write the 
effective probability that a generic edge resists failure as
\begin{equation}
p_\mathrm{eff}=p\cdot q^{\alpha},
\end{equation}
with $\alpha=\langle n\rangle\simeq 2.4$.  The system will then be resilient if 
$p_\mathrm{eff}=p_c$, which implies Eq.~(\ref{eq:pstar}).

Equation~(\ref{eq:pstar}) may be contrasted with a direct approximation of when 
an effective GSC forms.  From simulation results, we associate each transition 
with the value $p_\mathrm{mid}$, defined as halfway between $p_c$ and the 
lowest value of $p$ for which $\overline{l}_\varepsilon$ is maximal 
(\textit{i.e.}, the midpoint of the transition).

\section*{Acknowledgements}
RGM thanks the grant CEA-NRT `STARC' for financial support. MB is supported by 
the FET-Proactive project PLEXMATH (FP7-ICT-2011-8; grant number 317614) funded 
by the European Commission.

\section*{Author contributions}
Together, RGM and MB conceived of, and analysed the model, with RGM producing 
simulations.  The manuscript was written collaboratively between both authors.

\section*{Additional information}
The authors declare no competing financial interests.

\end{document}